\documentclass[letterpaper,12pt]{article}
\usepackage{tabularx} 
\usepackage{amsmath}  
\usepackage{graphicx} 
\graphicspath{ {./images/} }
\usepackage[margin=0.9in,letterpaper]{geometry} 
\usepackage[final]{hyperref} 

\usepackage[position=top]{subfig}

\usepackage[
  backend=bibtex, 
  style=numeric,
  natbib=true,
  sorting=nyt, 
  giveninits=true, 
  maxnames=999, 
  minnames=999, 
  date=year, 
]{biblatex}

\DeclareFieldFormat[article]{title}{#1} 

\DeclareNameAlias{default}{family-given}

\renewbibmacro*{author}{} 

\renewbibmacro*{issue+date}{%
  \printfield{issue}
  \setunit*{\addcomma\space}%
  \usebibmacro{date}
  \newunit}

\renewbibmacro*{issue+date}{%
  \printfield{issue}
  \newunit} 

\DeclareBibliographyDriver{article}{%
  \usebibmacro{begentry}
  \printfield{title}
  \newunit\newblock
  \usebibmacro{journal+issuetitle}
  \finentry}

\renewbibmacro*{begentry}{%
  \printnames{author}%
  \addspace\printdate\addperiod\addspace%
  \clearlist{date}
}

\usepackage[toc,page]{appendix}

\hypersetup{
	colorlinks=true,       
	linkcolor=blue,        
	citecolor=blue,        
	filecolor=magenta,     
	urlcolor=blue         
}

\begin{document}

\title{New method for subtraction of common background fluctuation for radio camera: Chunked Principal Component Analysis method}
\author{Pranshu Mandal, Tomu Nitta, Makoto Nagai, Nario Kuno}
\date{\today}
\maketitle

\begin{abstract}

We present a new algorithm, ChunkedPCA, to remove common background fluctuations from datasets acquired with a radio camera. ChunkedPCA is an improvement on using PCA to achieve fewer artifacts and better RMS on the cleaned dataset. The proposed algorithm determines the background fluctuation by grouping the detector pixels not used in the direct observation of the source. This group is then used to get the background fluctuation for that time and used to subtract the background from the data of all pixels. We apply ChunkedPCA for the numerical simulation data and a real observation data obtained with the MKID camera on the Nobeyama 45-m telescope to verify the effectiveness of the ChunkedPCA. We confirm that using the ChunkedPCA method preserves the flux of the astronomical sources and produces a cleaner baseline than the conventional PCA method.

\end{abstract}

\section{Introduction}

Subtraction of background fluctuation in an image is an integral part of the astronomical imaging pipeline which rectifies the intensity fluctuations due to atmospheric disturbances, electronics fluctuations, and other factors. The telescopes deployed around the world use some methods to get a flat background against the true signal. One of the most common methods used in radio astronomy for the subtraction of the fluctuation of background is the position-switching method, where the detector switches between a target source (On-source signal) and blank sky (Off-source signal). For the beam-switching method, the switching between On-source and off-source signals can be done much faster. Recently, radio cameras such as SCUBA2 (\hyperlink{cite.0@audley2004scuba}{Audley et al. 2004}), NIKA2 (\hyperlink{cite.0@adam2018nika2}{Adam et al. 2018}),  and BOLOCAM (\hyperlink{cite.0@mauskopf2000bolocam}{Mauskopf et al. 2000}), which can observe wide field-of-view (FOV), have been developed. The methods of the subtraction of the fluctuation of background for the cameras are the median filter method, and Principal Component Analysis (referred to as PCA from here on out) method (e.g. \hyperlink{cite.0@jenness1998removing}{Jenness et al. 1988}). Although the implementation of these methods is standard, (\hyperlink{cite.0@murtagh1987heck}{Murtagh et al. 1987}, \hyperlink{cite.0@hunziker2018pca}{Hunziker et al. 2018}), there are some poblems with both of them.  Both uses each time order data(TOD) to remove the fluctuation from itself which might introduce some artifacts in the cleaned data. The median filter method predicts the fluctuation quickly and fairly well. Although PCA is slower than the median filter, it is less sensitive to noise compared to median filter. PCA does this by extracting features that are common among all the TODs thus picking up common signals. However, in both methods when there is a strong source signal, the background is warped toward the signal peak. This may cause loss of signal intensity after the fluctuation subtraction (\hyperlink{cite.0@laurent2005bolocam}{Laurent et al. 2005}).

To solve the problem, we have developed Chunked PCA method. 
For radio camera with array detector, thanks to the large FOV, while some detector pixels observe the source, other pixels observe a blank sky, given the source is smaller than the FOV.
Our camera was installed on the Nobeyama 45-m telescope which can make use of On-The-Fly(OTF) (\hyperlink{cite.0@sawada2008fly}{Sawada, Tsuyoshi et al. 2008}) technique to make a raster scan of the sky. During the OTF observation, the pixels constantly scan the sky to collect TOD while being on and off source throughout the observation. In order to determine the background sky and the observed source intensities, we first divide the entire TOD into small chunks and then determine which pixels are off-source to apply the PCA method to only off-source data.

We verify the effectiveness of ChunkedPCA in getting a flat background by using simulated data as well as the data obtained with a Microwave Kinetic Inductance Detector (MKID) camera for the Nobeyama 45-m telescope which operates at 100 GHz band (\hyperlink{cite.0@nagai2018data}{Nagai et al. 2018}). It operates with 109 MKID detectors which simultaneously records a time series data from individual detector pixel.

This paper is organised as follows. We describe the conventional PCA algorithm in section 2 followed by a detailed explanation of the new ChunkedPCA algorithm in section 3. The algorithms are applied to the simulated and observation data in section 4 to demonstrate the advantages of the ChunkedPCA algorithm in getting a flat background. Finally, summarise our results in section 5.

\section{Principal Component Analysis(PCA)}

\begin{figure}[!htbp]%
    \centering
    \subfloat[\label{fig:given_data}]{{\includegraphics[width=.4\textwidth]{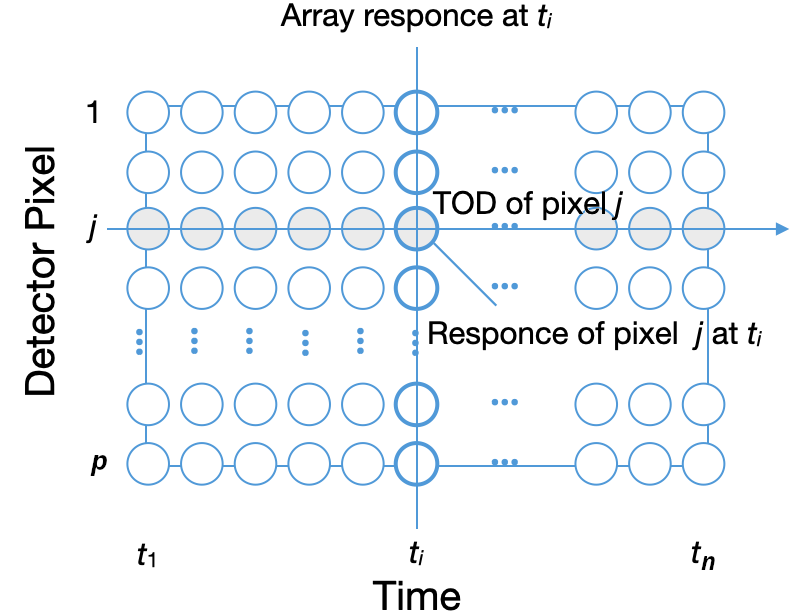} }}%
    \qquad
    \subfloat[\label{fig:flow_pca}]{{\includegraphics[width=.4\textwidth]{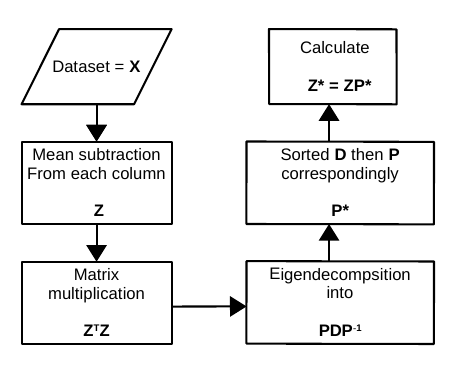} }}%

    \setlength{\belowcaptionskip}{-8pt}
    \caption{(a) The given data X. The i,j element of $X$ is written as $X_{i,j} = T_j(t_i)$, where $T_j$ is response of pixel $j$ and $t_i$ is $i^{th}$ sampling time. (b) Flowchart of the PCA algorithm.}%
    \label{fig:data_and_pca}%
\end{figure}

In order to make Chunked PCA easier to understand, we first explain conventional PCA. PCA is a method of transforming a correlated dataset of an observation into linearly uncorrelated variables (\hyperlink{cite.0@pearson1901liii}{Pearson, 1901}). The uncorrelated variables are then used on the observation to represent the dataset in reduced dimensions. We use this method to find the common mode from the data of all pixels of an array detector. The data from the detector is demonstrated in figure \ref{fig:given_data}, where the data contains $p$ number of TODs with $n$ number of data-points per TOD.

Since all the pixels record the sky intensity regardless of whether the source is in the beam of the pixel or not,
the recorded data always contains sky emission. This response keeps changing due to weather and atmospheric fluctuations during the observation. We refer to it as the sky fluctuations. PCA allows us to find such common fluctuations. Since it is observed by all the pixels during the observation, if we can find the common fluctuation in all the observed data, and subtract it, we can get a flat background.



First, we explain how the PCA algorithm works. In figure \ref{fig:flow_pca}, a flowchart of calculations in the PCA algorithm is presented. The steps are explained in detail below.

\textbf{- Step 1:} We start with a matrix \textbf{\textit{X}} which has \textit{n} rows and \textit{p} columns. For our dataset, it needs to be arranged such that the rows are time order data and columns are the different observations by individual detector pixels.

\begin{equation}
\textbf{\textit{X}} = 
\begin{bmatrix}
x_{1,1} & . & . & x_{1,p} \\
. & . & . & . \\
. & . & . & . \\
. & . & . & . \\
x_{n,1} & . & . & . \\
\end{bmatrix}^{n\times p}
\end{equation}

\textbf{- Step 2:} Then the mean of each column is subtracted from all the values in that column such that every column has a mean of 0. Let's call this centered matrix \textbf{\textit{Z}}. Then we calculate the covariance matrix, $\Sigma $ of \textbf{\textit{Z}} by finding \textbf{\textit{Z}}$^T$\textbf{\textit{Z}}. 

$$ \Sigma = \textbf{\textit{Z}}^T\textbf{\textit{Z}} $$

\textbf{- Step 3:} We use the $ \Sigma $ to find its eigenvectors and eigenvalues by the method of eigendecomposition. Thus $\Sigma$ can be represented as \textbf{\textit{PDP}}$^{-1}$, where \textbf{\textit{P}} is the matrix of eigenvectors and \textbf{\textit{D}} is the diagonal matrix with eigenvalues on the diagonal. Here, the first value of \textbf{\textit{D}}, $ \lambda_1$, corresponds to the first column of \textbf{\textit{P}}. We can always do the eigendecomposition since $\Sigma $ is a positive semidefinite matrix and symmetric.

\textbf{- Step 4:} Now, to give the result a proper weight, we sort the eigenvalues in the diagonal matrix \textbf{\textit{D}}, which are $\lambda_1, \lambda_2, \lambda_3 ..., \lambda_p$ and sort them by their values from largest to smallest. Also, we arrange the columns in \textbf{\textit{P}} according to the same arrangement. Let's call this sorted eigenvector matrix \textbf{\textit{P}}$^*$.




\textbf{- Step 5:} We can now calulate \textbf{\textit{Z}}$^* = $ \textbf{\textit{ZP}}$^*$. This \textbf{\textit{Z}}$^*$ contains columns where each observation is a combination of the original variables but the weights are determined by the eigenvectors. The eigenvalues are variations toward the eigenvector.




We have used a module in Scikit-learn  (\hyperlink{cite.0@scikit-learn}{Pedregosa et al. 2011}) called PCA to carry out all the calculations. It comes with ready to use open source algorithm that can give us the PCA analysis results with access to the eigenvalues and the eigenvectors. This algorithm gives us the principal components of the given dataset.

Now we multiply the eigenvector matrix with the original 2D dataset matrix \textbf{\textit{X}}. This gives us the dataset multiplied by the weight given by the eigenvectors. Assuming we take only the first 3 principal components, we add the weights for all three eigenvector rows over each pixel TOD and subtract it from the original TOD to get the final result as shown in equation \ref{eq:pca}, where $i$ is the datapoint of $j^{th}$ detector pixel, and $s$ is the number of principal components selected, $\mathrm{weight_k} * \mathrm{pcaVector}_{k,j}$ calculates the contribution from the $k^{th}$ principal component to the reconstructed value for pixel $j$.

\begin{equation}\label{eq:pca}
    \mathrm{TOD^{cleaned}}_{i,j} = \mathrm{TOD^{raw}}_{i,j} - \sum_{k=1}^{s} \mathrm{weight_k} * \mathrm{pcaVector}_{k,j}
\end{equation}


This is a [\textit{p} $\times$ \textit{n}] matrix where each row is an individual pixel and each column contains its corresponding TOD. This new TOD explains the variation in the original data with all the pixel's contributions included. This is regarded as the background which follows the atmospheric fluctuations appropriately. However, in trying to explain the variation among all the signals in the dataset, it can not help but give the signal features from astronomical objects a weight as well. These features resulting from the source signal create artifacts as shown in section 4. Hence, we propose a new algorithm for background fluctuation removal in the next section.

\section{Chunked PCA}

\begin{figure}[ht]%
    \centering
    \includegraphics[width = 10cm]{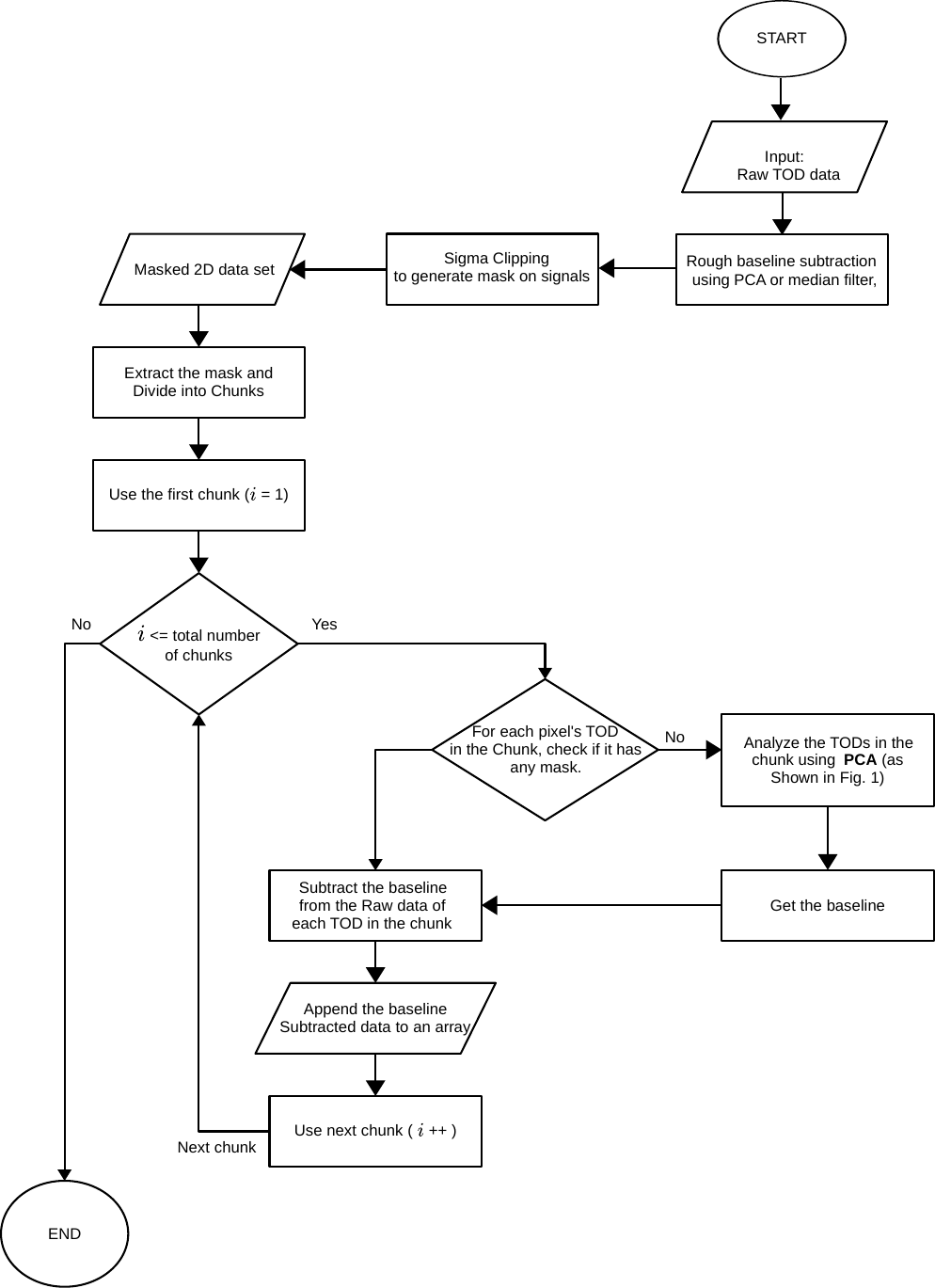}
    \caption{The flowchart diagram of the ChunkedPCA algorithm.}%
    \label{fig:flowchart}%
\end{figure}

We propose Chunked-PCA method to eliminate the above-mentioned artifacts from the background that comes due to the signal from a strong and compact source. We identify the detector that observed the source at any given time. The rest of the pixels that observed blank sky only recorded the intensity response of the sky fluctuation over time. These blank sky observations are used to derive their common mode as the average sky fluctuation. The common mode is then subtracted from the raw data to get a flat background. The algorithm is shown in figure \ref{fig:flowchart}.

We explain the algorithm in detail step by step.

\textbf{- Step 1:} We need to know when a particular pixel observed the source. There are two ways of going about it. The first is to have a table alongside the TOD that says when the pixel is on-source or off-source. This can be achieved by giving the source position in the sky and recording during observation as the source passes through each pixel's beam. It, however, is difficult to know the position of all sources in advance. The other way is to process the data later by some method that can find the intensity response rise as the source passes through the pixel's beam and flag it as such.

To do this, we apply sigma-clipping using \texttt{sigma-clip} from \texttt{astropy} module. Sigma clipping is a process of identifying the signal by setting a threshold in the sigma scale of the data (\hyperlink{cite.0@stetson1987daophot}{Stetson, 1987}). Any data above the threshold is clipped by masking it out. However, applying sigma clipping to data directly, will not work, since the data include large background fluctuation. Hence, we first do conventional PCA to remove the background fluctuation roughly. This does not need to be accurate, just well enough to flatten it so that the signals can be identified. Any other less computing intensive method such as median filter can also be used. After flattening the data, the peaks are identified using sigma-clipping. The mask contains the signals and hence, they can be used to identify when the source was in the beam of a pixel. This gives us a masked array of data where the mask is \texttt{True} for clipped values. Let's call this masked matrix \textbf{\textit{M}}.

\begin{figure}[ht]%
    \centering
    \includegraphics[width =7cm]{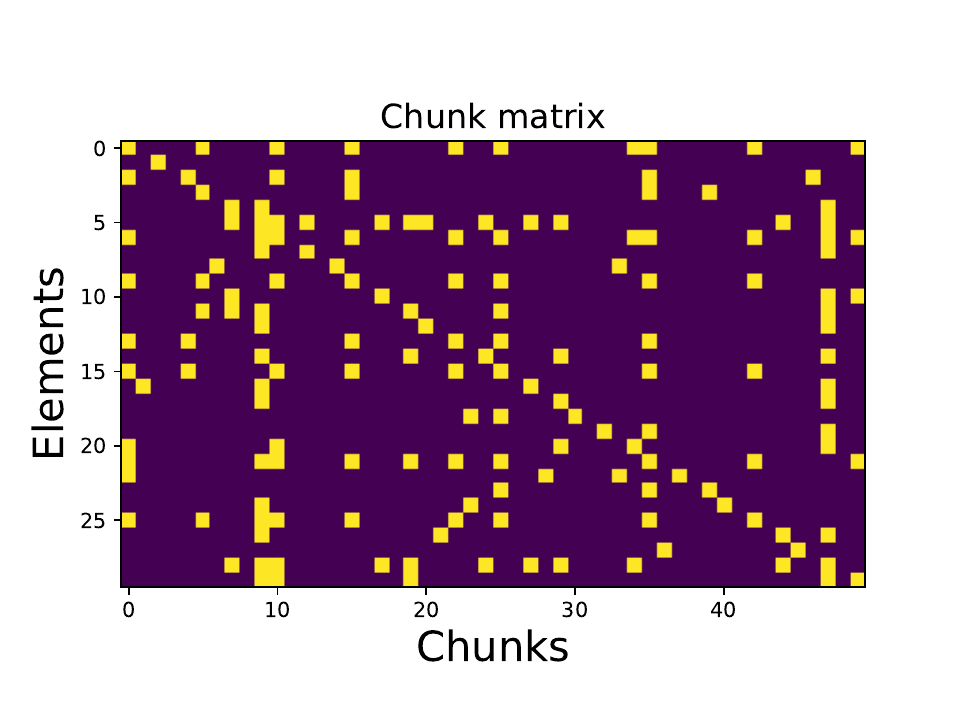}
    \caption{An example of a chunk matrix, where the yellow represents the presence of a mask, and the purple pixels mean that there are no masks, and the data is used for PCA. (This is a chunk matrix of 50 chunks)}%
    \label{fig:chunk_matrix}%
\end{figure}

\textbf{- Step 2:} Now we look for subsets in the data that are to be done PCA upon. Using the mask made from the previous step, we group some of the TOD together upto a certain length such that PCA can find the common mode in that group. This splits the entire observation into chunks of data that are separately analysed for a common mode. Then we look through every group of data and the corresponding values in \textbf{\textit{M}}. If it has any mask in it, we assign the pixel and its TOD in that chunk a \texttt{True} value. Since a better representation of this is to show it on a matrix, let's call it the chunk matrix \textbf{\textit{C}}. Figure \ref{fig:chunk_matrix} is a visual representation of this matrix which has a dimension of [number of pixels $\times$ number of chunk].

\textbf{- Step 3:} We perform the PCA one chunk at a time. The dataset \textbf{\textit{X}} used this time has a size of [number of pixels that does not have any mask $\times$ the length of the chunk TOD]. If the source is a point source, most of the pixels at any given time observe the blank sky. On the other hand, if a source is larger than the FOV of an array detector we use, we cannot use this method. Finally, the eigenvector found from each of these chunks is used to calculate the background of the corresponding participating pixels. However, the masked pixels do not have any eigenvector corresponding to their column. Hence, unlike the last method, we need to follow it up with an additional step to find the background for the masked pixels.

\textbf{- Step 4:} Finding the background for the masked pixels not included in the PCA analysis is performed as below. We can assume that the background fluctuation is similarly observed by all the pixels. Thus one way is to take the average background fluctuation observed by unmasked pixels as the background fluctuation for the masked pixels. An even safer method is to find the pixels that were not masked and are highly correlated with any given masked pixel. Then use its background as that for the masked pixel to eliminate any artifacts due to averaging blindly across all the pixels.

\textbf{- Step 5:} Finally, with all the chunks analysed, and every pixel with flat background, we stitch the data together. This gives us a cleaner background subtraction without any source signal artifacts.

\section{Application to simulation and observation data}

To test the performance of the ChunkedPCA we try to apply the method to simulation and observation data. Simulation data are randomly generated with a predefined common mode which we will try to remove. We shall then move on to real observation data of Mars obtained with an MKID camera installed on the Nobeyama 45-m telescope.

\subsection{Application to simulation Data}

\begin{figure}[h]%
    \centering
    \includegraphics[width =10cm]{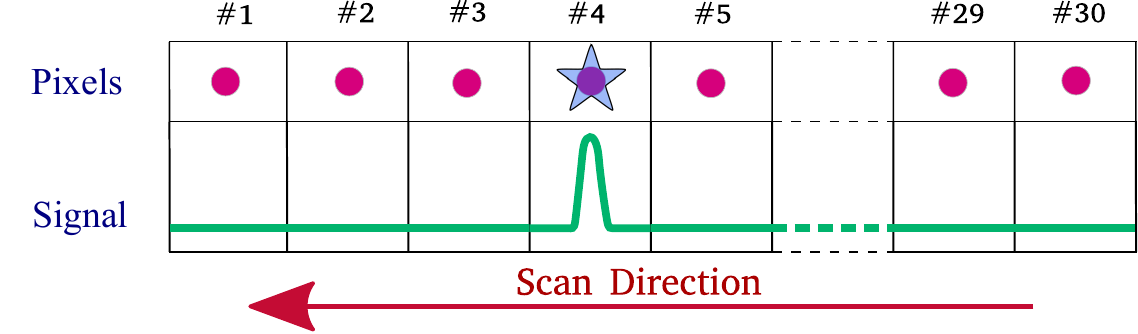}
    \caption{Illustration of a 30-detector pixel camera in spatial relation to the source in the sky, represented by a blue star. The pixel positions are shown as pink dots. As time progresses, the pixel array moves from right to left in the scan direction. At any given moment, only one pixel can observe the source. The signal panel shows the signal being received from the source by each pixel for a certain duration. This illustration shows the 4th pixel to be observing the source, hence it has a peak in its signal. The rest of the pixels are observing the blank sky.}%
    \label{fig:illustration}%
\end{figure}

The data corresponds to a raster scan with an array detector aligned regularly to the scan direction (as shown in fig \ref{fig:illustration}). For simplicity, we assume that the scan separation is the same as the beam separation which is larger than the beam size so that the source is detected only once by each detector pixel. We are going to simulate such an observation and use the data from the simulation for testing the algorithms.

\subsubsection{Simulation data}
\begin{figure}[h]%
    \centering
    \subfloat[\label{fig:common_mode}]{{\includegraphics[width=7.5cm]{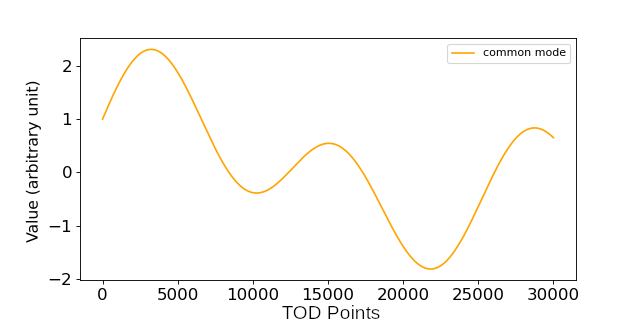} }}%
    \qquad
    \subfloat[.\label{fig:sample_data}]{{\includegraphics[width=7.5cm]{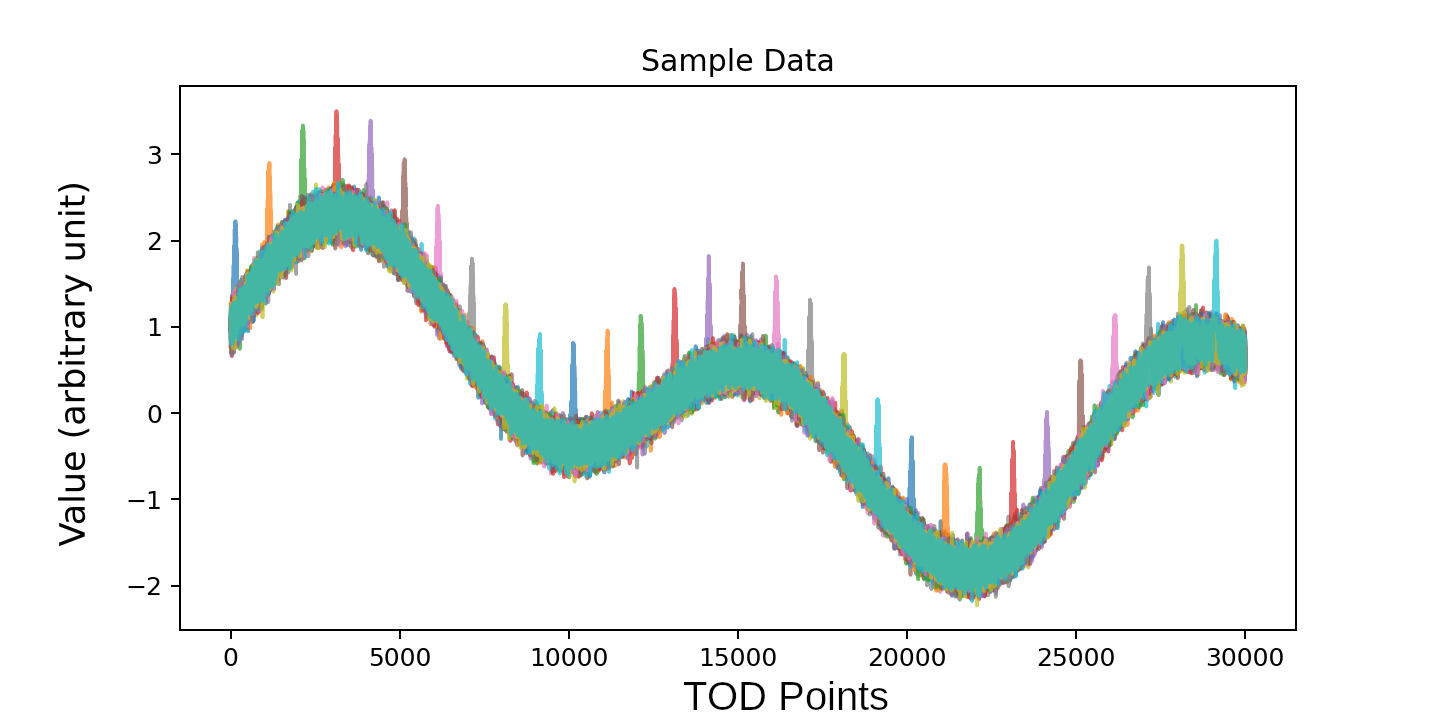} }}%
    \caption{Simulation data. (a) Background fluctuation which is seen by all pixels and varies slowly with time. (b) Simulated observation dataset consisting of 30 TODs, each with added Gaussian noise and a Gaussian peak simulating the source observation.}%
    \label{fig:data}%
\end{figure}

The simulation data consists of 30 Time Order Data(TOD) observed with the 30 detector pixels. The TOD consists of 30,000 recorded points in the form of columns.  The common mode, which is the consistently similar pattern seen in all the observations is an arbitrarily chosen combination of sine waves, given as $ \mathrm{sin}(t/6000) + \mathrm{cos}(t/5000) + \mathrm{sin}(t/2000) $, where $t$ is time(figure \ref{fig:common_mode}). To each of the TOD, we then add a Gaussian noise. To simulate observing a point source, we add a Gaussian peak in each TOD whose amplitude is 5 times larger than the rms of the Gaussian noise. The peaks are distributed along the time of observation. They are not overlapping each other, and maintain a uniform distance from the other peaks. Each TOD has just one such peak, as can be seen in Figure \ref{fig:sample_data}, where each color represents the TOD of a single detector.

\subsubsection{Baseline removal algorithms}

\begin{figure}[ht]%
    \centering
    \includegraphics[width = 7cm]{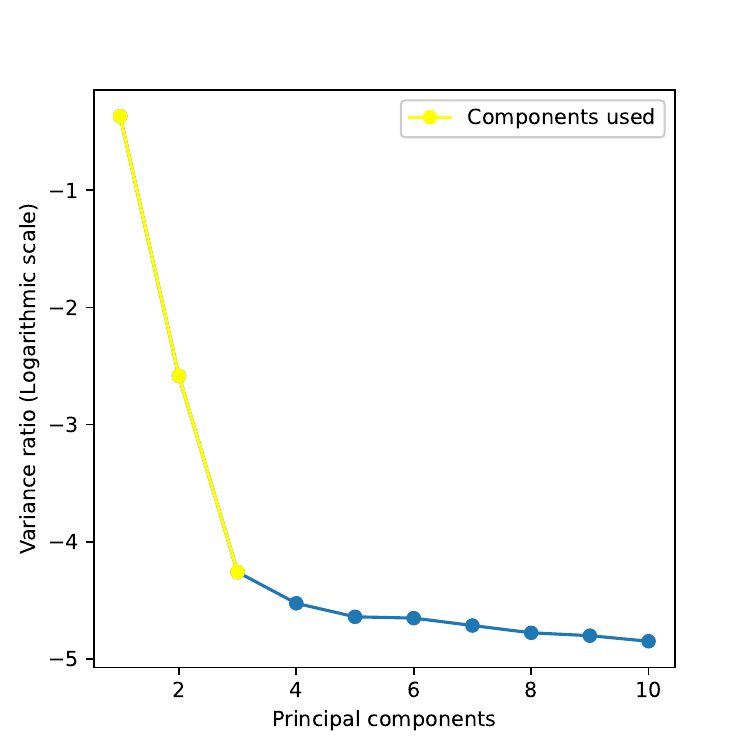}
    \caption{For the simulation data, the variance ratio following diminishing return as the number of components are increased for PCA.}%
    \label{fig:variance}%
\end{figure}

While implementing PCA, it is important to decide how many components to use to describe the data. For example, in the observation data, the first 3 principal components explain 78.15$\%$ of the variance in the data. Using more number of components does not add any meaningful variation to the data as shown in figure \ref{fig:variance}. Hence, we use 3 principal components, and therefore the eigenvector is a [3$\times$\textit{p}] matrix, where \textit{p} is the number of pixels in the camera. Each value in the eigenvector matrix represents the weight of each pixel used to get the corresponding principal component.

\begin{figure}[!htbp]%
    \centering
    \includegraphics[width =11cm]{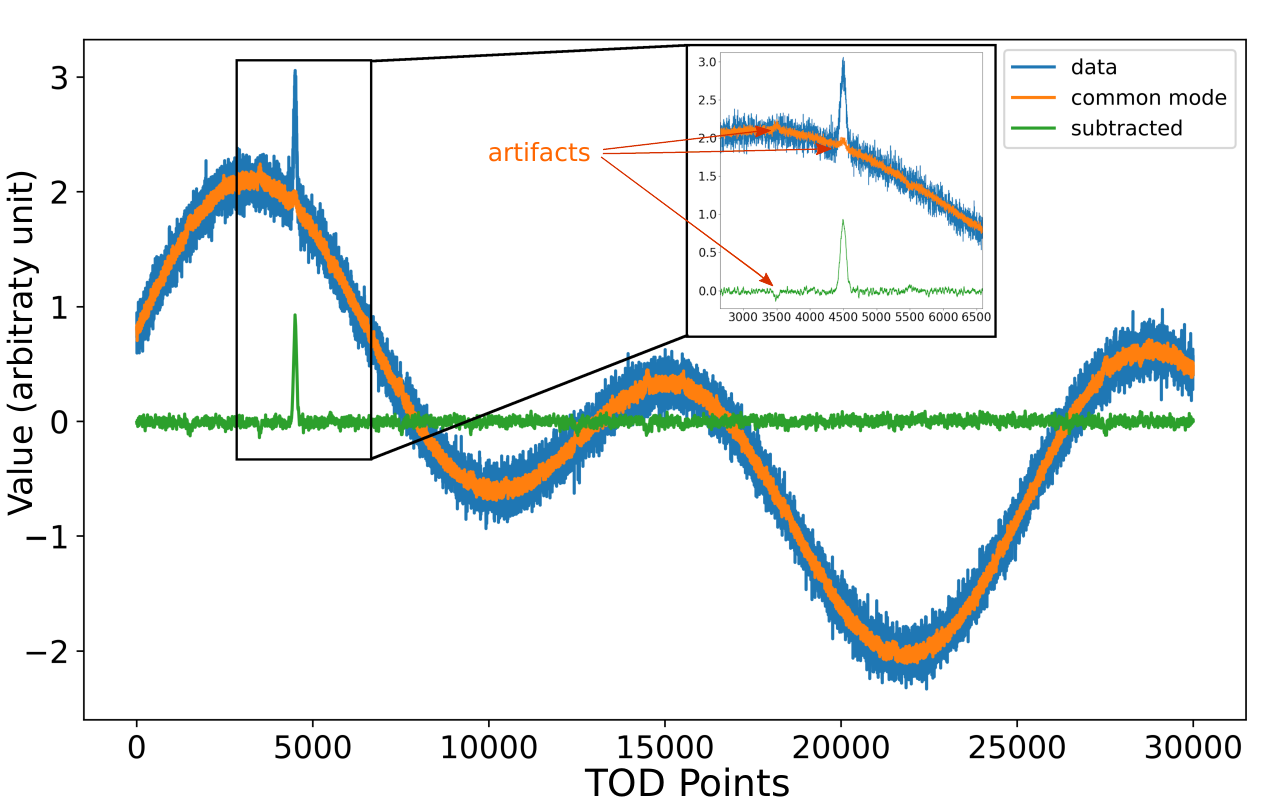}
    \caption{Results of conventional PCA for a single pixel. Its common mode, and the subtracted data are labeled on the top right. The zoomed-in part shows the distortions caused by the source signal in this pixel, as well as the neighboring data.}%
    \label{fig:pca_results}%
\end{figure}

Applying PCA over the dataset we find the common mode. Since the signals are not excluded, the PCA as expected generates an artifact as seen in figure \ref{fig:pca_results}. The zoomed part shows the artifacts in the common mode. When the common mode is subtracted from the data, the artifacts are carried over as well.


\begin{figure}[!htbp]%
    \centering
    \includegraphics[width =11cm]{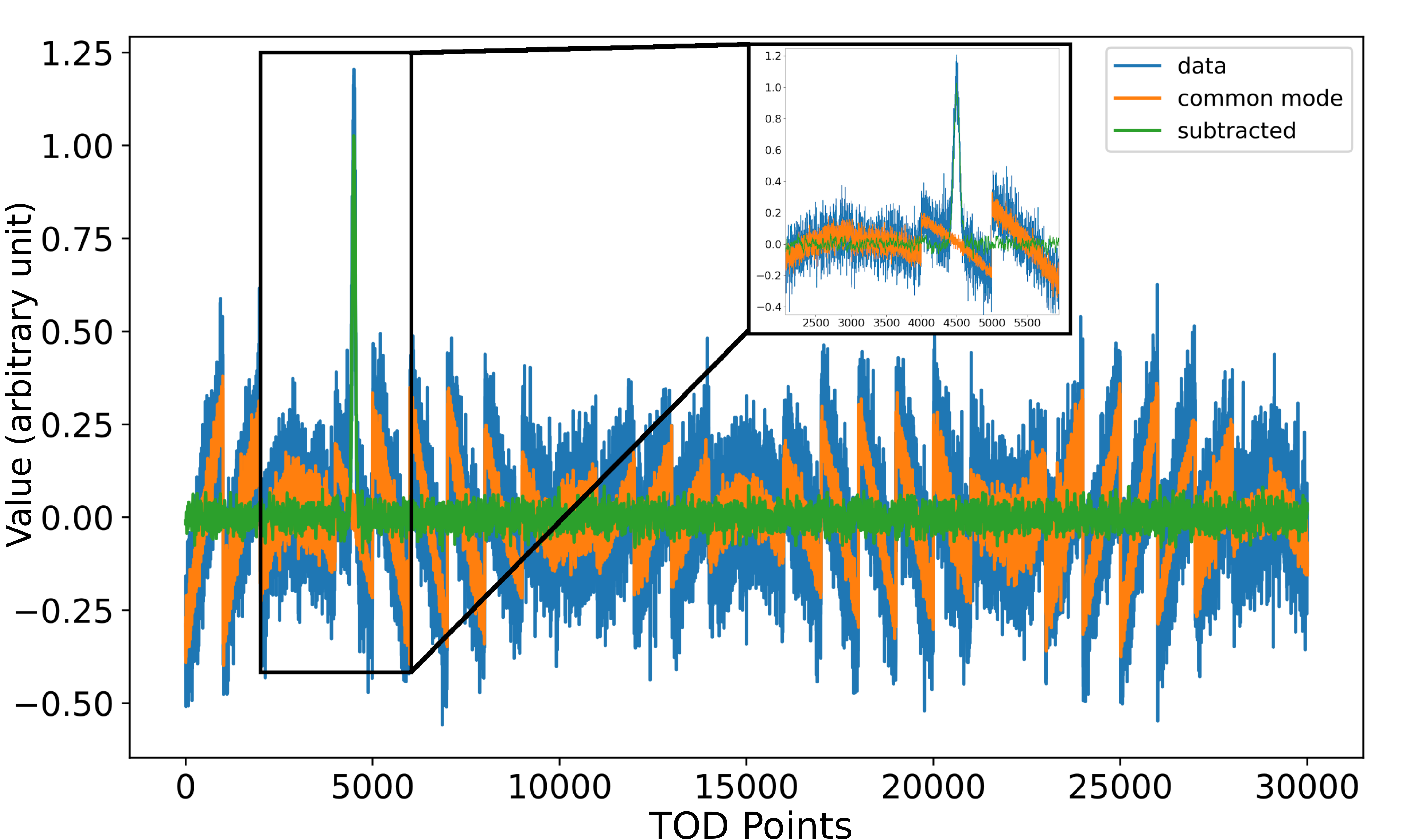}
    \caption{Result of ChunkedPCA for a single pixel. Its common mode, and the subtracted data are labeled on the top right. The zoomed-in part shows that the distortions caused by the source signal are not present.}%
    \label{fig:chunked_pca_results}%
\end{figure}

We also apply ChunkedPCA on the data and show the results in figure \ref{fig:chunked_pca_results}. It shows that the artifacts are not present in the common mode, and hence are not carried over to the subtracted values. Note that the data after ChunkedPCA do not resemble the common mode in Figure \ref{fig:chunked_pca_results}. This is due to the mean centering of the data which is required prior to PCA, where we subtract the mean of a given chunk of data from all of its values. This is helpful, as the subtracted data do not need any special attention to stitch together after performing the analysis by chunks.

Another popular method to flatten the baseline in such astronomical Radio time series data is Polynomial fitting. The process of obtaining a flat baseline is to do a 3rd or higher degree polynomial fitting to the mean of the TOD. Then the resulting fit is subtracted from the original TOD to obtain a flat baseline.

\subsubsection{Amplitude retention after baseline removal}

\begin{figure}[h]%
    \centering
    \includegraphics[width =0.6\textwidth]{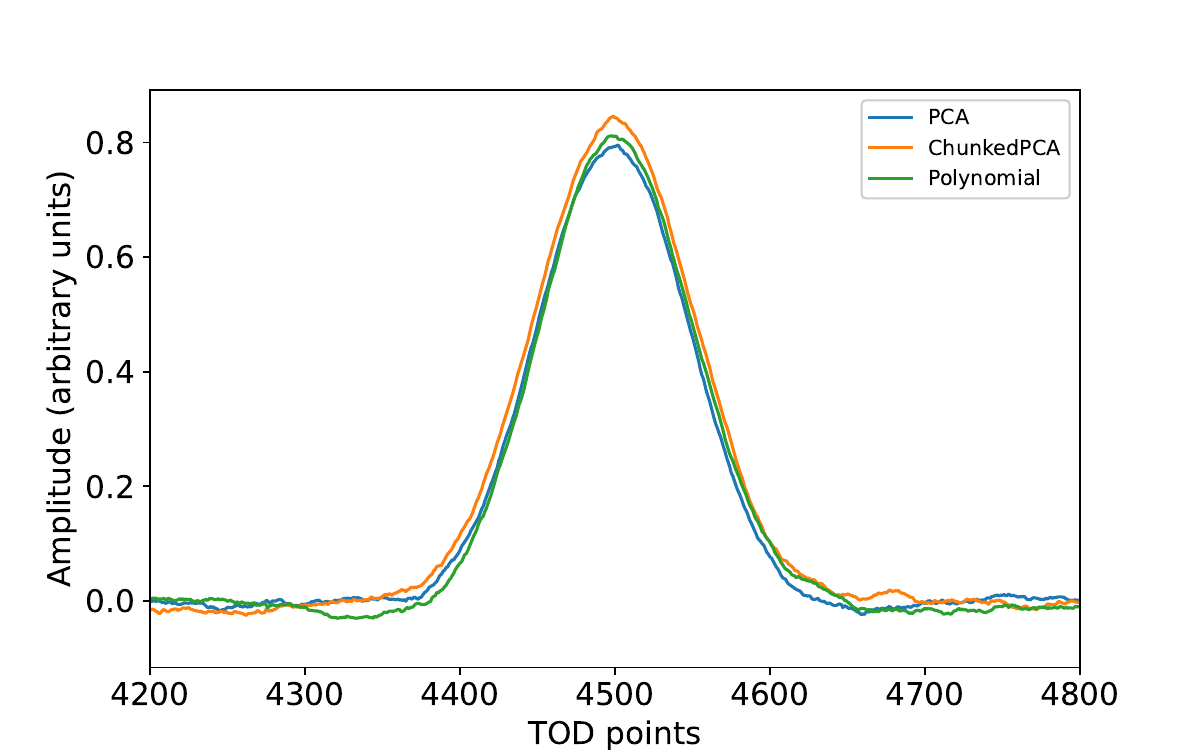}
    \caption{TOD of one of the detector pixel \#4 cropped to show the signal peak. Blue and green lines are the results of the conventional PCA and Polynomial fitting respectively. The orange line is the result obtained from ChunkedPCA.}%
    \label{fig:tod_subtracted}%
\end{figure}

By plotting the three cleaned signals obtained by the above methods gives us a flat baseline TOD to work with. As shown in figure \ref{fig:tod_subtracted} we can see the amplitude of the source signal after removing the baseline. It can be seen how Chunked PCA retains the original amplitude better than the other two methods. Thus by implementing the mask on signals, and then carrying out the common mode calculations using ChunkedPCA, we get a better amplitude retention after flattening the data.

\subsubsection{RMS comparison of cleaned sample data.}

\begin{figure}[!htbp]%
    \centering
    \includegraphics[width = 0.6\textwidth]{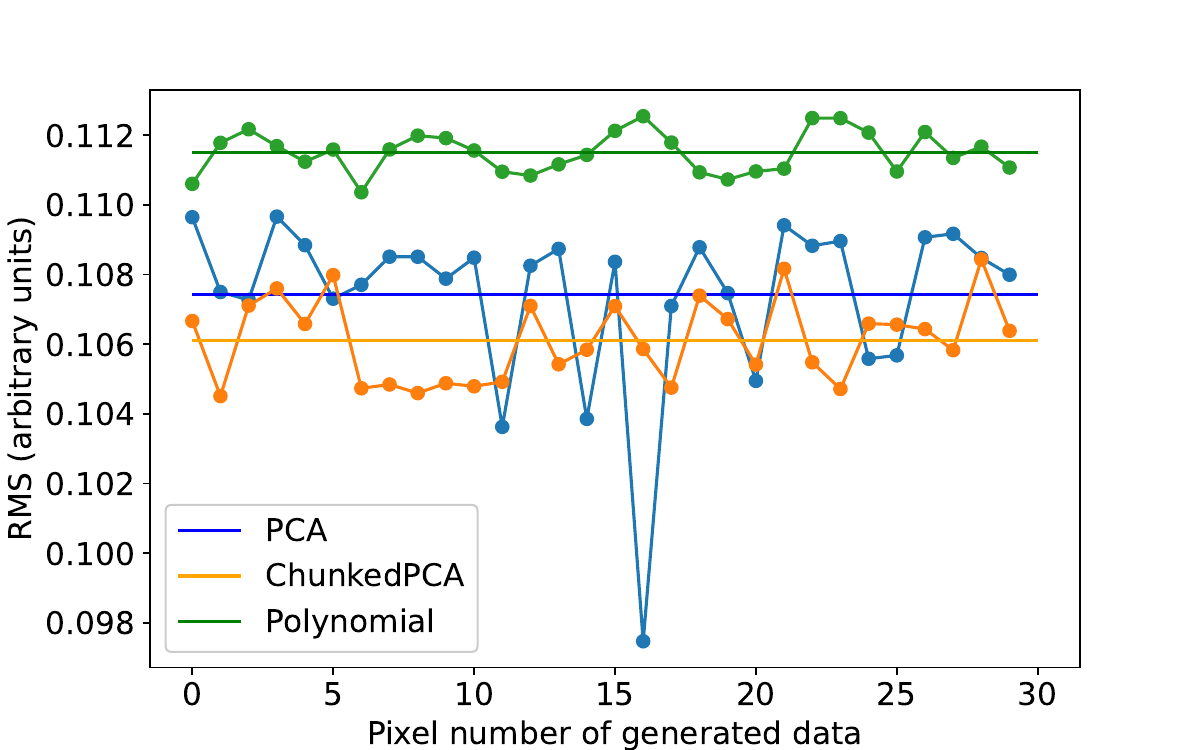}
    \caption{RMS value comparison after subtraction of sky fluctuations using PCA and ChunkedPCA.}%
    \label{fig:rms_single}%
\end{figure}

The noise removal using Polynomial, PCA and ChunkedPCA methods produces different levels of RMS in the cleaned signal as shown in table \ref{tab:mytable}. Polynomial being a rudimentatry fit to the mean of the TOD fails to follow the TOD as good as PCA based methods, resulting in a higher RMS value. PCA and ChunkedPCA is expected to be very similar in the unmasked regions. However since we expect the masked regions containing the signals to behave better using ChunkedPCA, we can expect an overall reduction in the RMS. This reduction is solely due to the removal of unwanted peaks in the baseline. We calculated the RMS  of all the 30 pixels of simulated data and we get the expected results as shown in figure \ref{fig:rms_single}. The average RMS is consistently lesser when using ChunkedPCA over PCA and Polynomial.

\begin{table}
    \centering
    \begin{tabular}{|l|l|} 
    \hline
    Method     & RMS value  \\ 
    \hline
    Polynomial & 0.1112     \\ 
    \hline
    PCA        & 0.1088     \\ 
    \hline
    ChunkedPCA & 0.1065     \\
    \hline
    \end{tabular}
    \caption{Algorithm used and the RMS values obtained after cleaning the data.}
    \label{tab:mytable}
\end{table}

\subsubsection{Reshaping the cleaned TOD}

\begin{figure}[!htbp]%
    \centering
    \subfloat[\label{fig:map_poly_15}]{{\includegraphics[width=.28\textwidth]{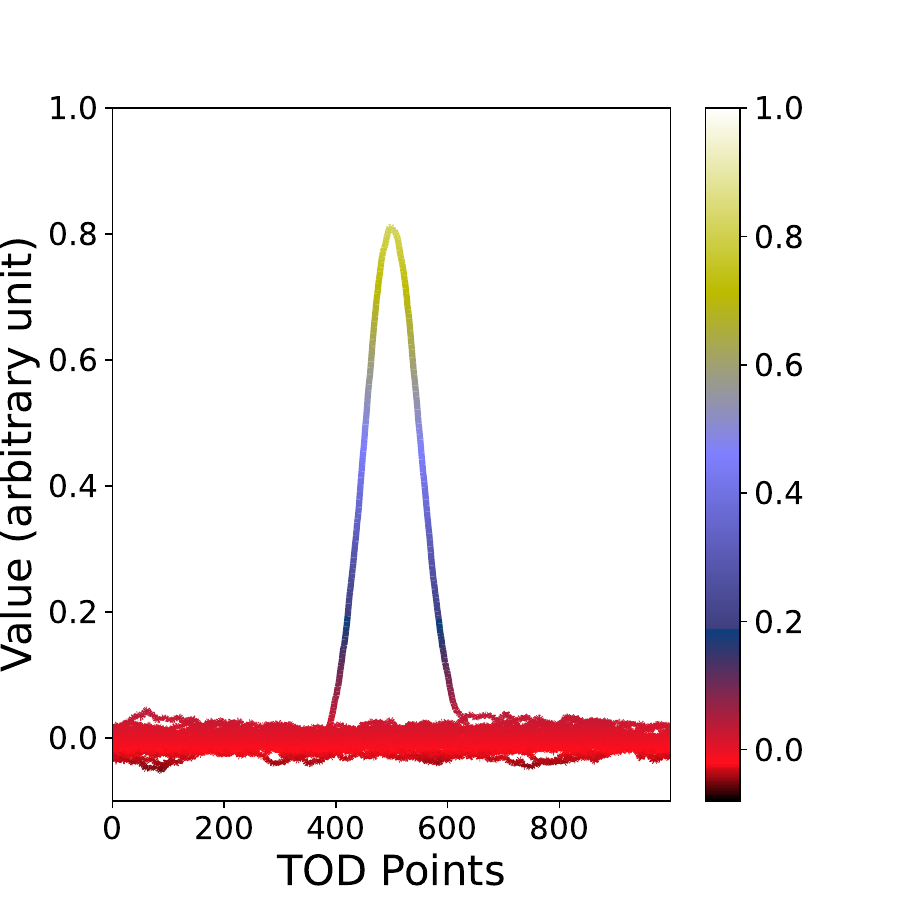} }}%
    \qquad
    \subfloat[\label{fig:map_db_15}]{{\includegraphics[width=.28\textwidth]{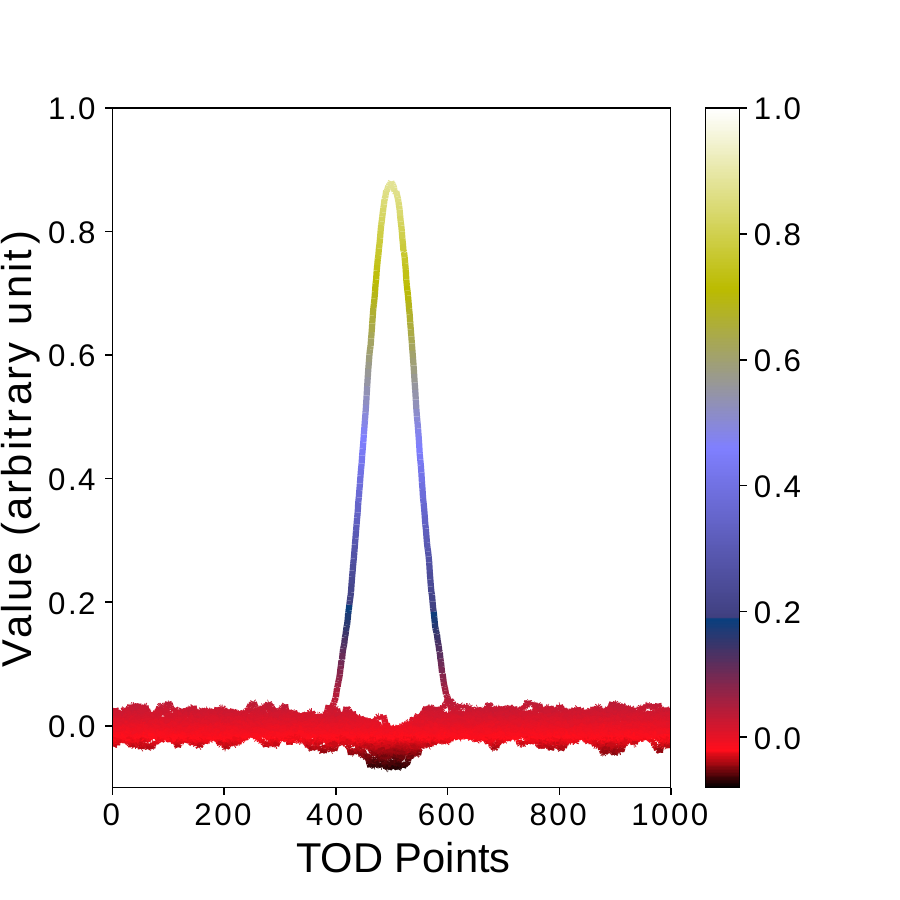} }}%
    \qquad
    \subfloat[\label{fig:map_15}]{{\includegraphics[width=.28\textwidth]{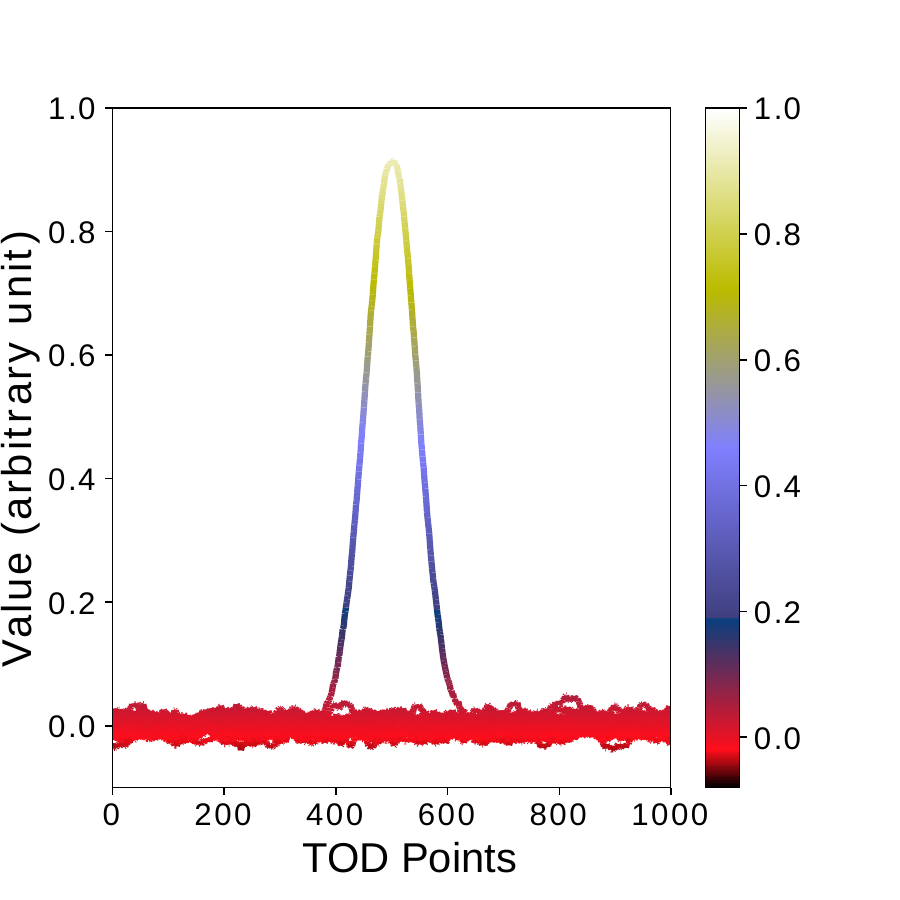} }}%

    \setlength{\belowcaptionskip}{-8pt}
    \caption{TOD of sample data in which only pixel \#4 observes the source. (a) Cleaned using Polynomial fitting, (b) Cleaned by PCA, (c) Cleaned by ChunkedPCA.} %
    \label{fig:image_plots}%
\end{figure}

To visualize and understand the efficiency of ChunkedPCA more clearly, we reshape the cleaned data into 30 segments and plot them as shown in Figure \ref{fig:image_plots}. Recall, we intentionally created the simulation data with 30 equally spaced signals. If we reshape the data that had 30,000 points to begin with, each line represents 1000 data points in the TOD of each pixel, where the signals, along with the artifacts they produce are lined up in the middle of the plot.

In figure \ref{fig:image_plots}, we have plotted the cleaned TOD in which the source is observed with the pixel \#4 using all three methods. As can be seen in the figure, when normal PCA is used, negative value artifacts show up in the baseline of other pixels. Polynomial does not give rise to such artifacts, but the cleaned baseline is much noisy, and does not follow the true baseline as well as the other two methods. Using ChunkedPCA effectively removes the artifacts while following the baseline more consistently.

\subsubsection{Effect of different amplitude in the signal.}

\begin{figure}[h]%
    \centering
    \includegraphics[width =10cm]{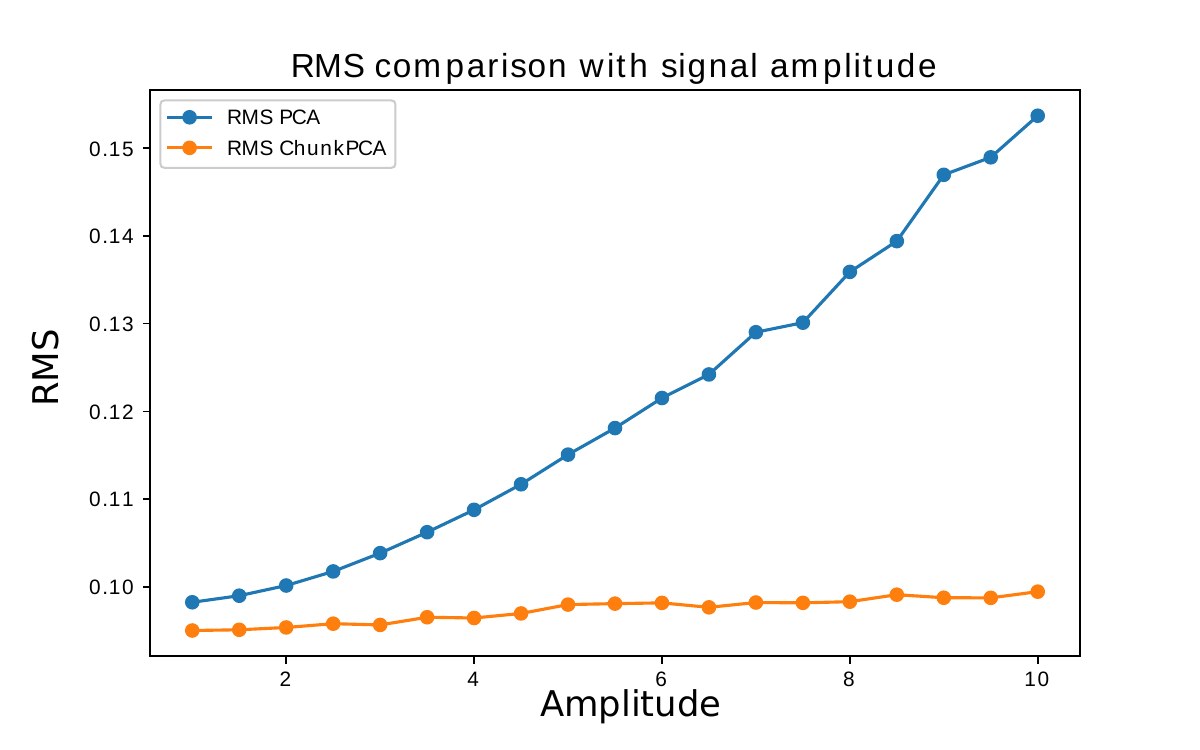}
    \caption{As the amplitude of the signal peak increases, the RMS stays relatively similar in ChunkedPCA. }%
    \label{fig:rms_amp_comp}%
\end{figure}

As we have discussed earlier, the amplitude of the signal peak in the data causes the artifact in the predicted baseline using PCA. We expect the bigger the amplitude, the bigger the artifact it causes. ChunkedPCA should take care of this artifact no matter how big the amplitude is. We plot the RMS difference, using different amplitude of the source, and the combined image of all the samples to see the effect in figure \ref{fig:rms_amp_comp}. We see the expected rise in the rms of conventional PCA, because the more substantial peaks in one of the observations is strong enough to skew the baseline towards it. However, since ChunkedPCA masks the peaks out, the baseline is independent of the amplitude.

\subsubsection{Effect of chunk size on ChunkedPCA.}

A key parameter in the entire algorithm of ChunkedPCA is the chunk size. It is the size of datapoints used to do the PCA calculation. In general, we know that the minimum number of datapoints must be equal to or above the number of principal components we aim to reduce the dimensionality to. However, there is no upper-limit on the number of datapoints in each chunk as such. It must be noted that there are some rules that need to be followed in order to get a good result. We shall discuss them here.

In ChunkedPCA, the number of TOD available to calculate the baseline   per chunk is a big factor. If the chunk size is large, it may cause one chunk to contain the source signal in multiple pixels’ TODs. This will cause the pixels used to calculate the blank sky fewer. This will raise the noise in the baseline calculated. For example, if the number of chunks is 3, the chunk size is one-third the length of the entire TOD. Considering we are dealing with our simulated data, where the intervals  between source signals are equal for different pixels observing them. The first chunk will have one-third of the TODs masked. That is, the first 10 TODs will be masked out, and only the rest 20 pixels TODs will be used to calculate the baseline for the chunk. It is not ideal to lose so many TODs for blank sky calculation because the lesser the number of TODs used to find the common mode, the higher the noise in the baseline.

\subsection{Application to observation data.}

We present the effects of the two methods, PCA and ChunkedPCA, on a real observation data where the background fluctuation removal is necessary. The data shown here is observation of Mars obtained with an MKID camera installed on the Nobeyama 45-m telescope for observations in the 100GHz band. The camera has 109 pixels all of which record the signal in its field-of-view simultaneously. Observations from all the pixels collectively form the dataset we are going to use.

The data was collected during a raster scan of a 4'$\times$4' patch of sky with Mars at its center. Out of the 109 pixels in the detector, 82 of them produced good data with a beamsize of 20" on average.

\begin{figure}[!htbp]%
    \centering
    \subfloat[\label{fig:pca_beam_b} 
                    ]{{\includegraphics[width=6cm]{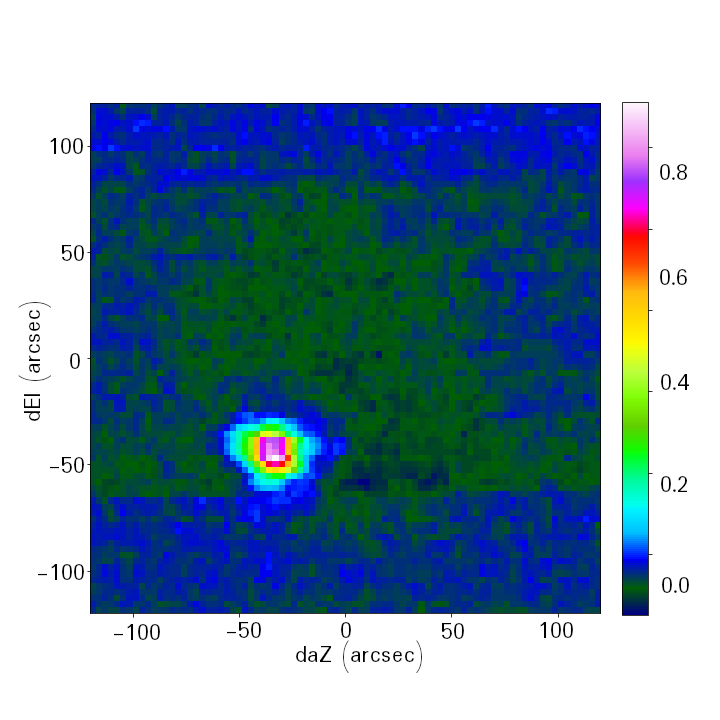} }}%
    \qquad
    \subfloat[\label{fig:chunk_beam} 
                     ]{{\includegraphics[width=6cm]{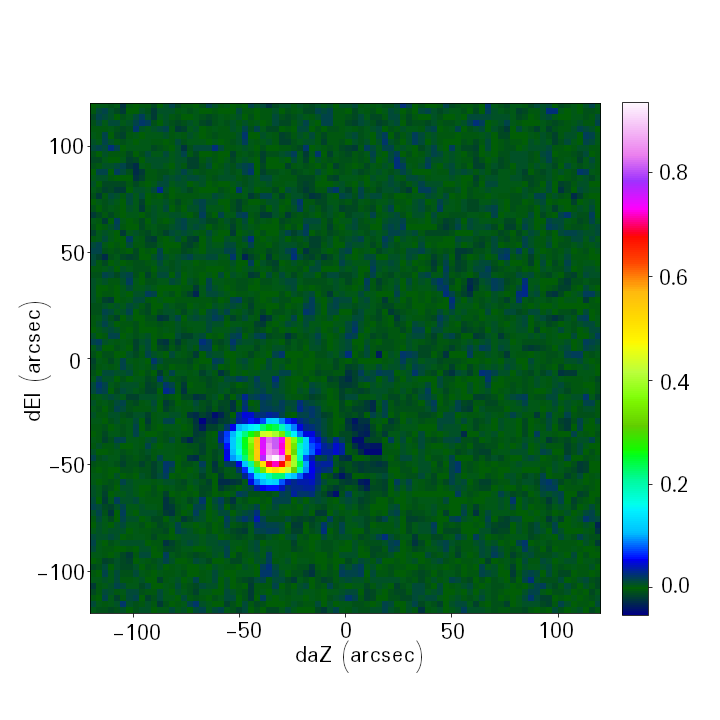} }}%
    
    \caption{Beam pattern of one of the pixels of the Nobeyama 45m MKID camera obtained by observing Mars. a) Map made by PCA. b) Map made by ChunkedPCA.}%
    \label{fig:pca_beam}%
\end{figure}

We need to keep in mind that the arrangement of the pixels in the detector causes it to observe the celestial source at a different time during the raster scan of the observation as shown in the simulation data. If the source size is relatively large compared to the FOV of the array detector, a large number of pixels might be observing the source, and just a few left observing the sky. In that case, it should be pointed out that as more and more pixels are excluded from finding the eigenvectors, the eigenvector might not explain the data very well if the variation is very large and the number of pixels used is very low resulting in a poor baseline.  But, for our Mars observation, with the large FOV($\approx$ 3') of our array detector with beamsize of 20" for each pixel, that is not a problem.  

\begin{figure}[h]%
    \centering
    \subfloat[\label{fig:pca_cleaned}]{{\includegraphics[width=6cm]{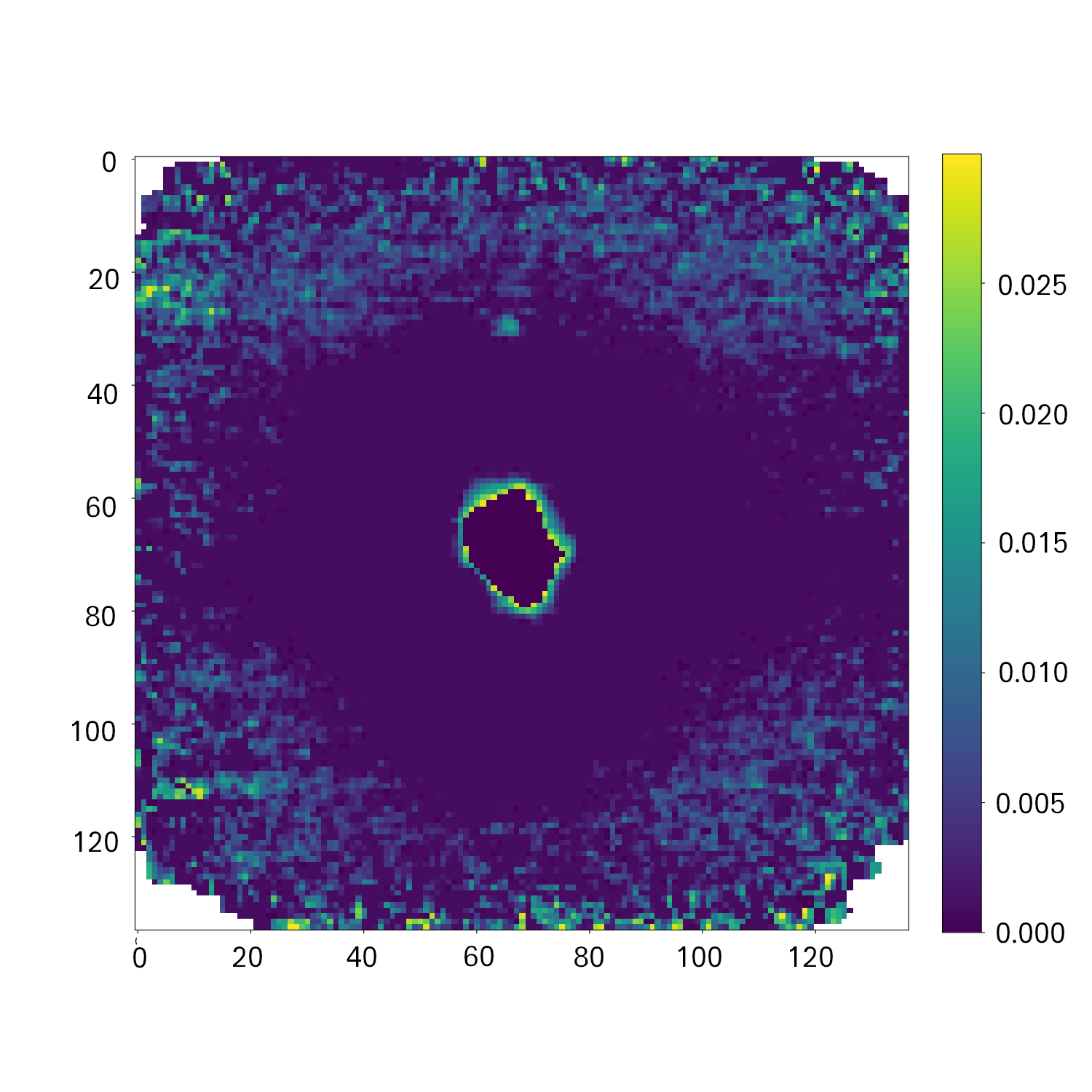} }}%
    \qquad
    \subfloat[\label{fig:chunk_cleaned}]{{\includegraphics[width=6cm]{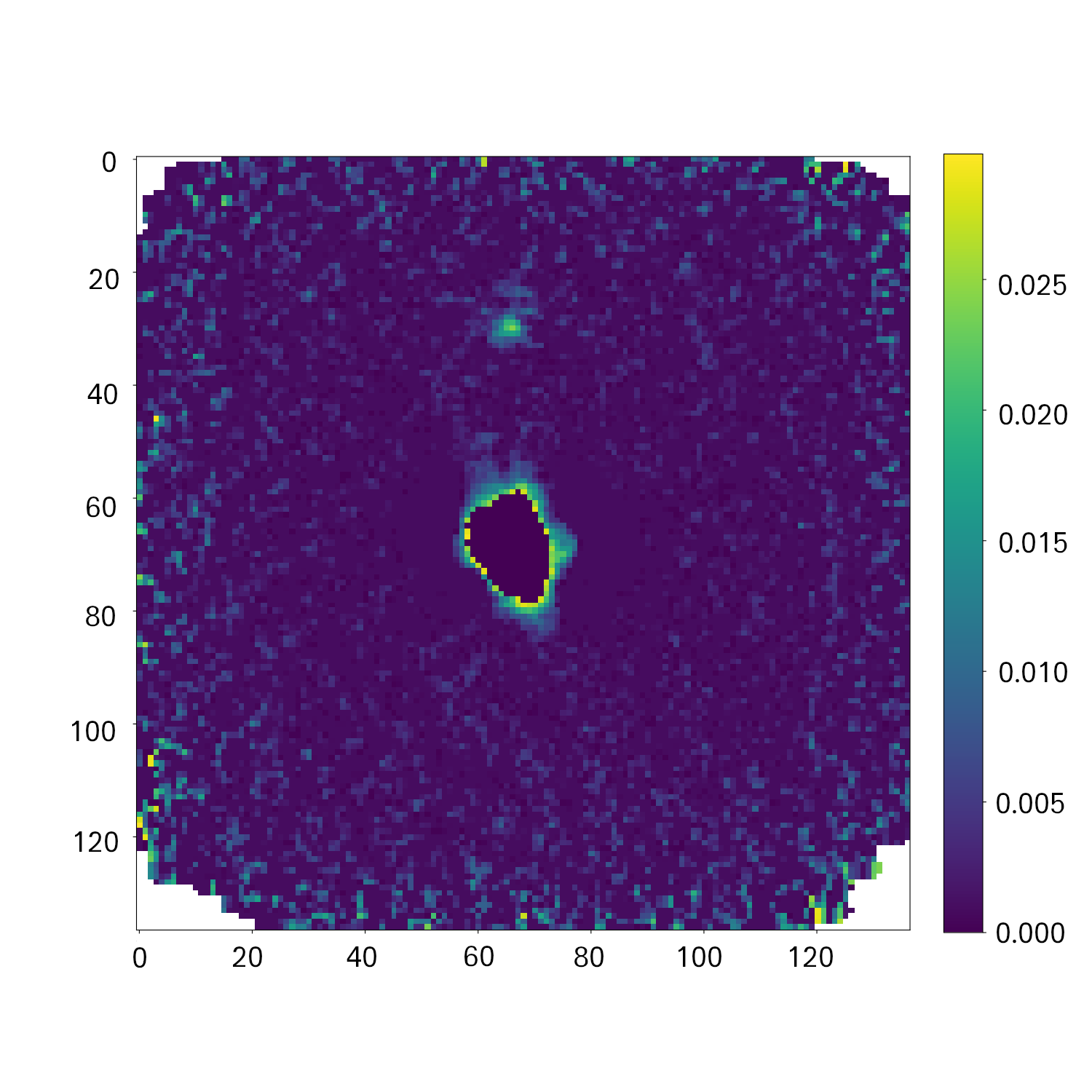}}}%
    \caption{A comparison to show the difference in the two algorithms. a) Cleaned by conventional PCA, RMS =0.0050, b) Cleaned by ChunkedPCA, RMS = 0.0036.}%
    \label{fig:rms_pca_comparison}%
\end{figure}

To summarize, We apply the baseline subtraction method to the observed TOD. Then the cleaned TOD is used to create 2D maps for each TOD using the module HEALPix (\hyperlink{cite.0@gorski2005healpix}{Gorski, 2005}).The results of the two algorithms are compared in figure \ref{fig:pca_beam} where a single pixel is used to create the map. The images show the effectiveness of ChunkedPCA in cleaning the background. Another comparison is shown in figure \ref{fig:rms_pca_comparison} where the combined map from data of all the pixels is created, and then it is clipped at -15dB value of the peak. The RMS of all the pixel values in the image is then calculated to give us a quantitative comparison between the images. This is the RMS value of the background noise. The circular shape around the figure \ref{fig:pca_cleaned} is a result of the artifacts caused by conventional PCA. The size of the artifact is approximately the field-of-view of the MKID camera(3').

\begin{figure}[h]
    \centering
    \includegraphics[width=7cm]{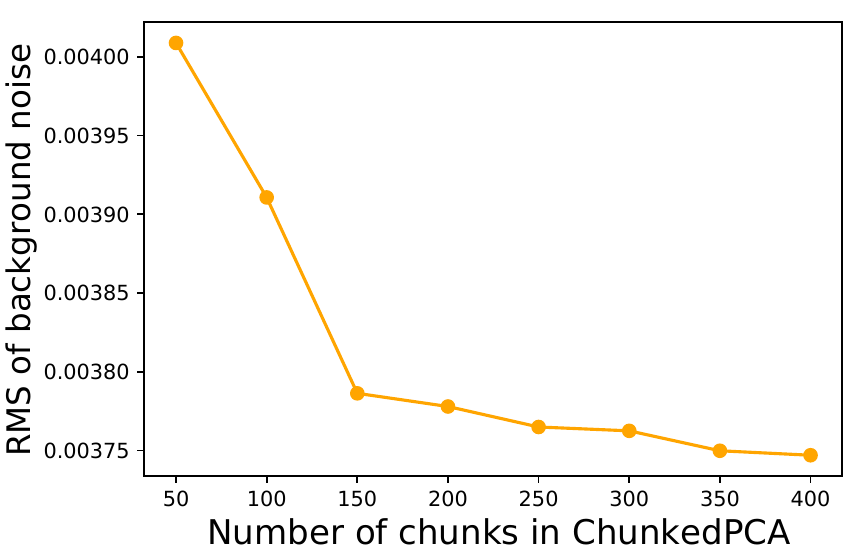}
    \caption{Dependence of RMS on the number of chunks. As we increase the number of chunks, the background noise rms keeps decreasing.}
    \label{fig:num_chunk_rms}
\end{figure}

We can see the effectiveness of the ChunkedPCA algorithm on our combined image. When using ChunkedPCA, the background noise RMS is lesser than the RMS of the map generated by using the conventional PCA method. We have further investigated the effect of the chunk-size variable of the ChunkedPCA algorithm. We notice that the smaller the chunk-size the lesser the RMS value as shown in figure \ref{fig:num_chunk_rms}. However, there is a lower limit to the chunk-size. It cannot be smaller than the total number of TODs used in a technical way.

\begin{figure}[h]
    \centering
    \includegraphics[width=9cm]{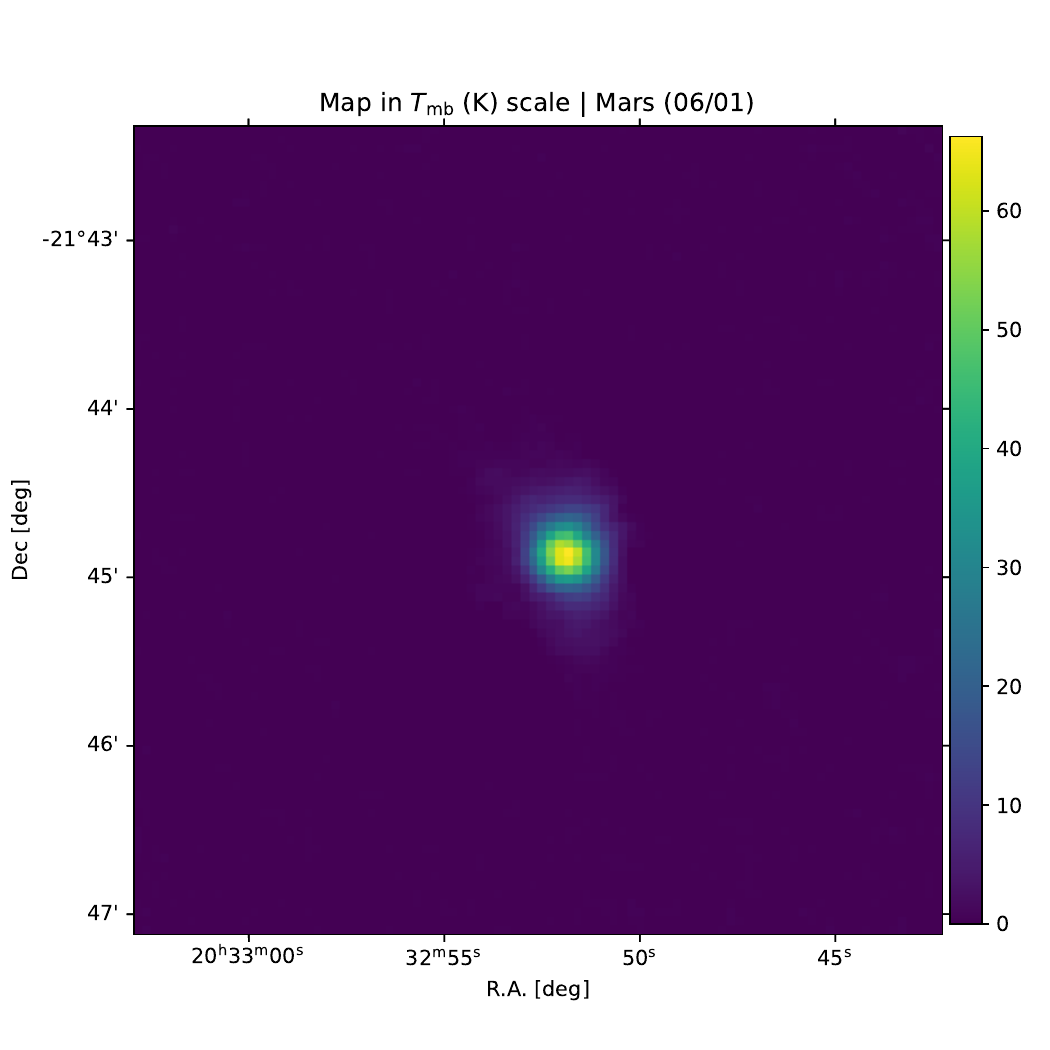}
    \caption{Final synthesized Mars image after background subtraction and scaled to main-beam Temperature.}
    \label{fig:mars_map_final}
\end{figure}

It is also important to note that the data needs to be scaled to proper units of measurement. To be properly scaled the raw TOD needs to go through 3 stages, $0^{th}$ order baseline subtraction, higher order baseline subtraction and then finally scaling. Our dataset is acquired using Chopper-wheel method which can be decomposed into offset subtraction and chopper-wheel scaling. The offset subtraction is the $0^{th}$ order baseline subtraction. Since ChunkedPCA preserves the data scaling, the chopper-wheel scaling and other kinds of scaling can be applied before and also after the higher order baseline subtraction using ChunkedPCA. After the baseline is subtracted, we have scaled the data to main-beam Temperature($T_{mb}$). The final map of the Mars observation is presented in figure \ref{fig:mars_map_final}.

\newpage 
\section{Conclusion}

We demonstrated a new algorithm for subtraction of background fluctuation that can decorrelate the readout data and provide a clean baseline. It is designed specifically for detectors with a large field-of-view. The philosophy behind the method is to utilize the large field-of-view of the camera to distinguish between off-source and on-source signals and then, define the background fluctuations using the off-source signal only.

It is developed on the basis of the PCA method and is a very lightweight algorithm that is fast and can be implemented using any regular computer. We have tested the algorithm on simulation as well as observed data and demonstrated the improvements it brings over conventional PCA. We presented how the strength of the signal is preserved even if the amplitude of the source signal is high. Using our new method, we get lower rms of the noise in the cleaned baseline. The cleaned time series data is also used to synthesize images to demonstrate the effectiveness of removing artifacts. The images also show the quality of the cleaned baseline is much flatter when ChunkedPCA is applied.

\section{Data Availability}
The data underlying this article will be shared on reasonable request to the corresponding author. The code used in this work is available publicly, and can be found in the (\hyperlink{cite.0@Chunked_PCA}{ChunkedPCA}) repository.






\nocite{*}

\printbibliography
\end{document}